\documentclass[aps,pre,twocolumn,showpacs]{revtex4}
\usepackage{graphicx}
\usepackage{subfigure}

\begin{document}
\title{A design principle for improved 3D AC electro-osmotic pumps}
\author{Damian Burch}
\author{Martin Z. Bazant}
\email{bazant@math.mit.edu}
\affiliation{Institute for Soldier Nanotechnologies and Department of
Mathematics, Massachusetts Institute of Technology, Cambridge, MA 02139-4307,
USA}
\date{January 12, 2008}
\pacs{47.57.jd, 47.61.Fg, 82.47.Wx, 82.45.-h}
%47.57.jd 	Electrokinetic effects
%47.61.Fg 	Flows in micro-electromechanical systems (MEMS) and nano-electromechanical systems (NEMS)
%82.47.Wx 	Electrochemical engineering
%82.45.-h 	Electrochemistry and electrophoresis

\begin{abstract}
Three-dimensional (3D) AC electro-osmotic (ACEO) pumps have recently been
developed that are much faster and more robust than previous planar designs.
The basic idea is to create a ``fluid conveyor belt'' by placing opposing ACEO
slip velocities at different heights.  Current designs involve electrodes with
electroplated steps, whose heights have been optimized in simulations and
experiments.  Here, we consider changing the boundary conditions---rather than
the geometry---and predict that flow rates can be further doubled by
fabricating 3D features with non-polarizable materials.  This amplifies the
fluid conveyor belt by removing opposing flows on the vertical surfaces,
and it increases the slip velocities which drive the flow.
\end{abstract}

\maketitle

\section{Introduction}
Microfluidic pumps are crucial components of lab-on-a-chip systems.  There is
growing interest in exploiting various phenomena of induced-charge
electro-osmosis (ICEO)~\cite{iceo2004a,iceo2004b} due to the lack of moving
parts, favorable scaling with miniaturization, tunable local flow control, and
low operating voltage suitable for portable or implantable devices.  The most
advanced technology of this type is based on AC electro-osmosis
(ACEO) around micro-electrodes.  ACEO was discovered by Ramos {\it et
al.\/}~\cite{green2000b, gonzalez2000, green2002}, who described steady,
time-averaged vortices over a pair of identical, co-planar electrodes applying
an AC voltage.  Ajdari~\cite{ajdari2000} predicted that breaking spatial
symmetry would generally lead to directional flows, and thus to microfluidic
pumps.  Based on this principle, the first ACEO pumps were built using
arrays of asymmetric pairs of planar electrodes, according to the design of
Brown, Smith, and Rennie~\cite{brown2001, studer2004}.

Motivated by ICEO flows around three-dimensional (3D) metal
structures~\cite{iceo2004a,iceo2004b,levitan2005}, Bazant and
Ben~\cite{bazant2006} recently predicted that the flow rate of ACEO pumps can
be increased by more than an order of magnitude (at the same voltage and
minimum feature size) by creating a ``fluid conveyor belt'' with arrays of
non-planar electrodes.  This was validated experimentally by Urbanski
{\it et al.\/}~\cite{urbanski2006} using gold electrodes with electroplated
steps.  Current work has optimized the step height in
simulations~\cite{olesen_thesis} and experiments~\cite{urbanski2007} for this
particular class of designs.

In this Rapid Communication, we predict that by modifying the boundary
conditions in addition to the geometry, the fluid conveyor belt can be further
amplified and the driving slip velocities increased.  We begin by explaining
the design principle in simple terms, building on the arguments of Bazant and
Ben~\cite{bazant2006}.  We then use the same simulation methods to validate the
theory and predict improved robustness and doubling of the flow rate compared
to current designs.

\section{Grooved Designs}
\begin{figure}
     \centering
     \subfigure[]{
          \label{fig:cartoon_flat_E}
          \includegraphics{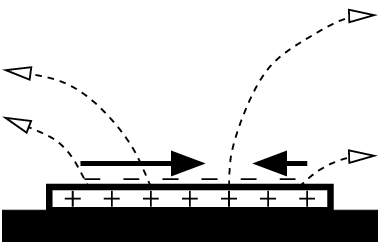}}
     \subfigure[]{
          \label{fig:cartoon_flat_U}
          \includegraphics{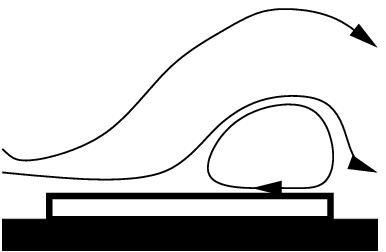}}
     \subfigure[]{
          \label{fig:cartoon_plated_E}
          \includegraphics{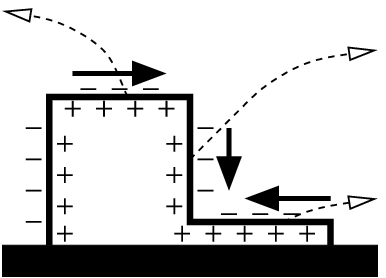}}
     \subfigure[]{
          \label{fig:cartoon_plated_U}
          \includegraphics{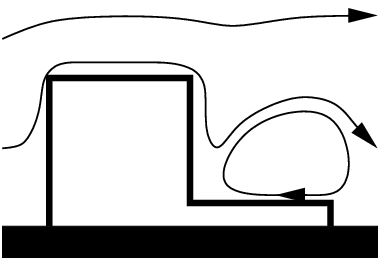}}
     \subfigure[]{
          \label{fig:cartoon_grooved_E}
          \includegraphics{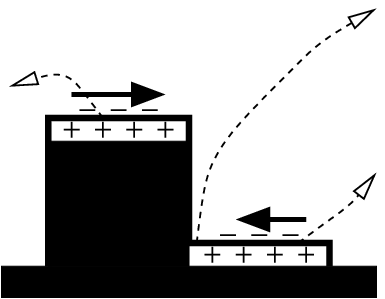}}
     \subfigure[]{
          \label{fig:cartoon_grooved_U}
          \includegraphics{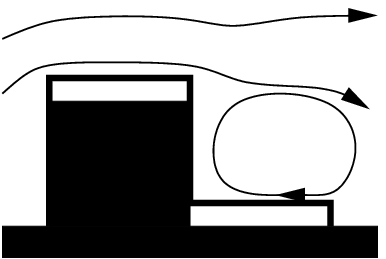}}
        \caption{ Design principles for ACEO pumps, shown in 2D cross-section.
	  Left: typical in-phase electric field lines (dashed lines) and slip
	  velocities (thick arrows) around one electrode in a periodic
          array.  Right: time-averaged streamlines (solid curves).  Top
          row (a,b): one electrode in an asymmetric planar array,
          showing the competition between opposing ACEO flows.  Middle
          row (c,d): the existing ``plated'' 3D ACEO design, which
          creates a fast fluid conveyor belt.  Bottom row (e,f): our
          faster ``grooved'' design, where the electrode is broken
          into two flat, electrically-connected parts, thereby removing
          counterproductive ACEO slip on the side wall.}
     \label{fig:cartoon}
\end{figure}

The motivation behind the discovery of 3D ACEO was to remove the competition
between different slip velocities along electrode surfaces, as shown in
Fig.~\ref{fig:cartoon}.  ACEO generally drives electro-osmotic slip in
opposite directions along different sections of each electrode, and Ajdari's
idea is to bias this competition so that one direction ``wins''.  In planar
pumps with asymmetric electrodes, symmetry is broken by making one electrode
and one inter-electrode gap in each pair wider than the other, thus generating
slightly more slip in one direction.  Though directionality is achieved,
a portion of each electrode is ``counteracting'' in the sense that its surface
slip is working against the net pumping (see Figures
\ref{fig:cartoon_flat_E} and \ref{fig:cartoon_flat_U}).

This competition is turned into cooperation in 3D ACEO pumps by raising the
productive part of each electrode relative to the counteracting part.  The
vortices above the counteracting portions are then recessed relative to the
bulk fluid.  These are ``rollers'' in the fluid conveyor belt:  their tops are
moving in the direction of the pumping and are vertically aligned with the
productive part of the electrode.

There is a problem with this design, however.  Because the sides of the steps
are polarizable, the adjacent fluid will have a significant double layer, and
there will be a vertical electro-osmotic slip (see
Fig.~\ref{fig:cartoon_plated_E}).  This slip acts \emph{against} the sides of
the rollers, slowing them down and reducing the effectiveness of the net
pumping (see Fig.~\ref{fig:cartoon_plated_U}).

In the spirit of~\cite{bazant2006}, we can improve performance by minimizing
this new source of competition.  In principle, we could eliminate the vertical
slip altogether by making the sides of the steps (but not the tops) relatively
non-polarizable.  This should then lead to a more effective fluid conveyor belt
and better overall performance (see Figures \ref{fig:cartoon_grooved_E} and
\ref{fig:cartoon_grooved_U}).

This can be achieved experimentally, for instance, by building the entire step
out of a non-polarizable material and then depositing thin electrodes atop the
step and in the recess.  Alternatively, one might etch the recesses into the
substrate rather than building the steps up from it.  Regardless of the actual
fabrication process, we shall henceforth refer to any system whose steps have
non-polarizable sides as the ``grooved'' design, and the case in which the
sides of the steps are polarizable as the ``plated'' design.  Note that
a grooved design was originally used to illustrate the fluid-conveyor-belt
principle in the context of a fixed-potential ICEO pump~\cite{bazant2006}, but
it has never been applied to 3D ACEO pumps, in simulations or experiments.  All
work has focused on plated designs, and the competition between vertical slip
velocities has not been recognized.

We have seen how grooved designs offer greater efficiency in the sense that
there is less counteractivity in the fluid flow.  However, the grooved designs
also offer greater forcing in the form of higher average slip velocities along
the tops of the raised parts of the electrodes.  This results from stronger
electric fields near the top-left edges of the electrodes, which is precisely
where most of the slip is generated.

To explain these differences in the electric fields,
we refer again to Figures \ref{fig:cartoon_plated_E} and
\ref{fig:cartoon_grooved_E}.  In the grooved design, the shortest distance
between two electrodes is between the right edge of the recessed part of the
left electrode and the left edge of the raised part of the right electrode.  In
the plated design, the shortest distance is much smaller as the bottom of the
vertical side of the right electrode is itself part of the electrode.  However,
as a consequence, the \emph{important} electric field lines in the plated
design are longer:  those starting at the left edge of the raised part of the
right electrode must reach much further into the left electrode before
terminating.  See Fig.~\ref{fig:Efield} for more complete,
numerically-generated plots.

Thus the important electric fields in the grooved design are stronger because
their field lines are shorter.  This has another consequence for the system:
the time-scale for ACEO flows is $\tau=\lambda L/D$ \cite{bazant2004}, where
$\lambda$ is the Debye screening length and $D$ is an ionic diffusion constant.
Typically, $L$ is taken to be a characteristic length scale of the electrode
geometry; however, it is really the length scale appropriate for changes in
electric potential.  Therefore the grooved design will have a shorter characteristic time scale, and so will operate at higher frequencies.

\section{Numerical Simulations}
Numerical simulations allow us to test the grooved design and explore its
consequences.  We employ the standard low-voltage model from previous studies
(e.g.~\cite{levitan2005,bazant2006}), using the same codebase as
in~\cite{urbanski2007}.  As noted above, experiments have validated the use of
this model to predict qualitative trends in 3D
ACEO~\cite{urbanski2006,urbanski2007}, at least at low voltage and low salt
concentration.  To focus on the impact of boundary conditions on the side
walls, we study only the simplest grooved geometry in which the electrode
half-widths and inter-electrode gaps all have the same length.  This is not
a drastic restriction since we still achieve pumping velocities which are
within a few percent of those generated by the optimal designs found by Olesen
using the same model~\cite{olesen_thesis}.

%\vfill\eject
%\begin{figure*}
%     \centering
%     \subfigure[]{
%             \includegraphics{Efield_plated.eps}}
%     \subfigure[]{
%             \includegraphics{streamlines_plated.eps}}
%     \subfigure[]{
%             \includegraphics{Efield_grooved.eps}}
%     \subfigure[]{
%             \includegraphics{streamlines_grooved.eps}}
%     \caption{ caption }
%\end{figure*}
%\vfill\eject

\begin{figure}
     \centering
     \subfigure[]{
	     \includegraphics{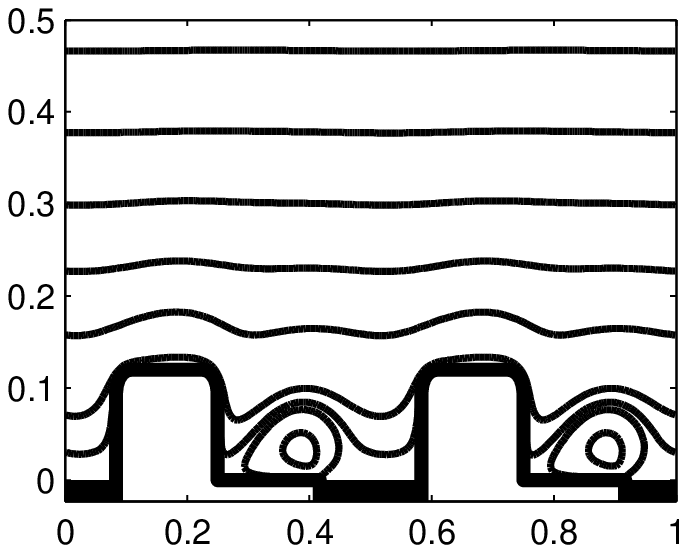}}
     \subfigure[]{
	     \includegraphics{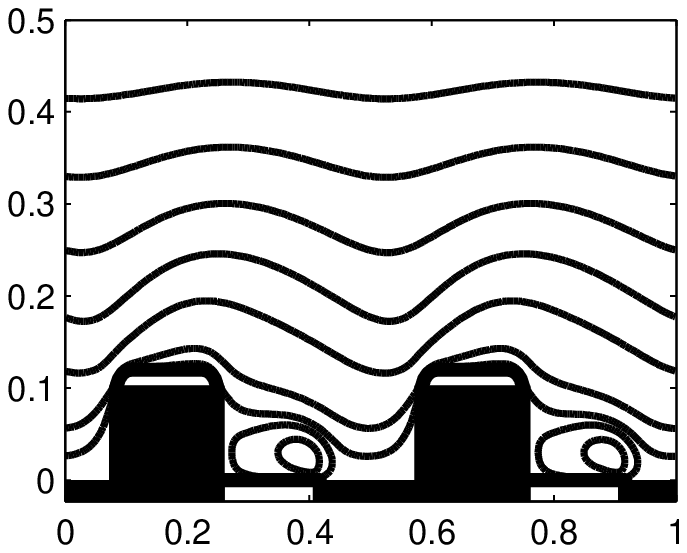}}
           \caption{ Simulated streamlines of the time-averaged 3D
             ACEO flow around one electrode pair in a plated (a)
             and grooved (b) design of the same geometry,
             demonstrating the principle in Fig.~\ref{fig:cartoon}.
	     Each case was simulated at its respective optimal
	     frequency (c.f.\ Fig.~\ref{fig:performance_frequency}).
	     Only 2D cross-sections are shown.  The electrode corners have
	     been rounded to improve convergence. }
     \label{fig:streamlines}
\end{figure}

\begin{figure}
     \centering
     \subfigure[]{
	     \includegraphics{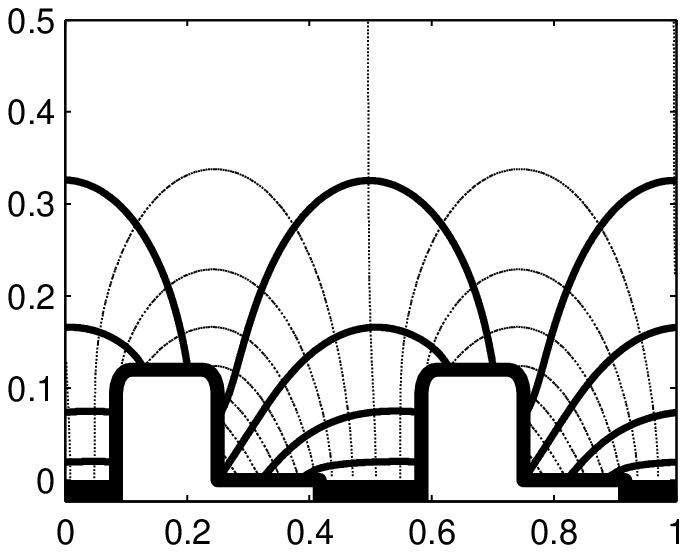}}
     \subfigure[]{
	     \includegraphics{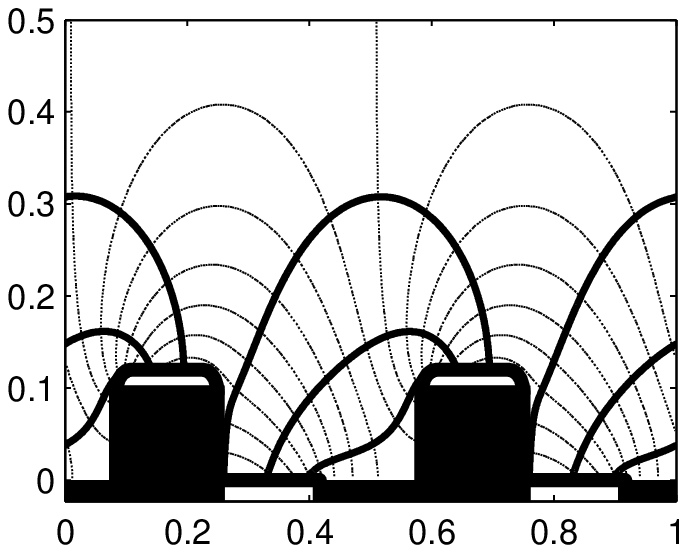}}
	   \caption{ Equally-spaced contours of the in-phase electric potential
	   (thin, dotted lines) and representative electric field lines (thick,
	   solid lines) for the simulations described in
	   Fig.~\ref{fig:streamlines}. }
     \label{fig:Efield}
\end{figure}

The computed streamlines for one particular case are shown in
Fig.~\ref{fig:streamlines}.  The results are as predicted:  the grooved design
exhibits higher slip velocities, and it doesn't suffer the strong ``dip'' seen
in the streamlines of the plated case.  This is in agreement with Figures
\ref{fig:cartoon_plated_U} and \ref{fig:cartoon_grooved_U}.

\begin{figure*}
     \centering
     \subfigure[]{
             \label{fig:performance_frequency}
	     \includegraphics{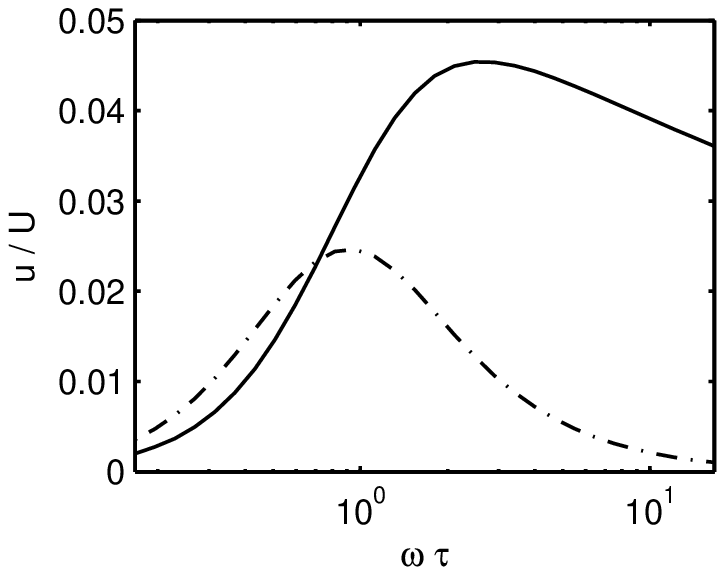}}
     \subfigure[]{
             \label{fig:performance_stepheight}
	     \includegraphics{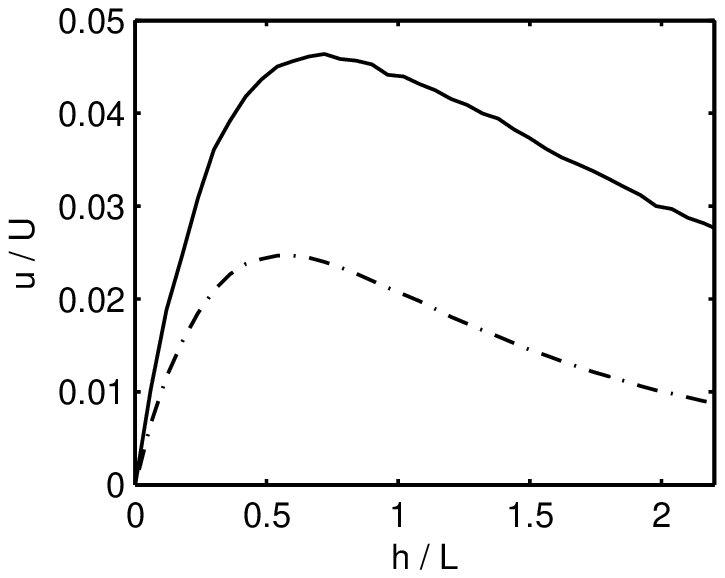}}
     \caption{The dependence of the simulated average pumping velocity on
	   AC frequency $\omega$ (a) and step height $h$ (b) for the grooved
	   (solid curves) and plated (dotted curves) designs shown in
	   Fig.~\ref{fig:streamlines}.  We follow standard
	   practice~\cite{bazant2006,urbanski2007} in non-dimensionalizing
	   lengths using the minimum electrode gap width $L$, frequencies using
	   the inverse $1/\tau$ of the RC time defined earlier, and velocities
	   using $U=\epsilon V^2/\eta L$ ($\epsilon$ is the solution's
	   permittivity, $V$ is the peak electrode voltage, $\eta$ is the
	   dynamic viscosity). }
     \label{fig:performance}
\end{figure*}

Because moving from polarizable to non-polarizable step sides offers greater
forcing and improved efficiency, it results in significantly faster pumping
velocities.  Indeed, for any step height, the linear model predicts that the
grooved design outperforms the plated design by at least 50\% (see
Fig.~\ref{fig:performance_stepheight}).  Moreover, near the optimal step
heights, the grooved design has 60\% stronger forcing and 20\% better
efficiency (defined to equal the average pumping velocity divided by the
forcing), yielding pumping velocities which are almost twice as fast as the
comparable plated design.  This is not as dramatic as the difference between
stepped and flat geometries, but it is still quite significant.

Figure \ref{fig:performance} shows the results of two brief, parametric
studies.  They clearly demonstrate that across a wide range of step
heights and driving frequencies, the grooved design is significantly faster.
Moreover, as predicted above, we see that the ideal operating frequencies are
higher in the grooved case.  Indeed, when the frequency is plotted on
a logarithmic scale (Fig.~\ref{fig:performance_frequency}), the shorter grooved
time scale causes the grooved velocity curve to be shifted to the right.  This
results in a crossover frequency below which the plated design is faster.
However, this occurs well below the optimal frequencies, so is unlikely to be
important in practice.

Figure~\ref{fig:performance_frequency} also reveals another important advantage of the grooved design:  its performance curve is much less sharply peaked around its optimal operating frequency than that of the plated design.  This can be explained in terms of the efficiency defined above.  At high frequencies, only the double layers near the edges of the electrodes have time to charge, so these will be the only places with significant slip velocities.  The left edge of the raised portions of the electrodes will still generate more positive slip in the grooved design because, as described above, the corresponding electric field lines are shorter.  However, the same reasoning (and Fig.~\ref{fig:Efield}) allows us to conclude that the electric field lines emanating from the right edge of the recessed parts of the electrodes will be \emph{longer} in the grooved case, leading to weaker negative slip.  The bigger negative slips and smaller positive slips in the plated design lead to much bigger vortices which reach well above the electrodes.  This impedes the bulk fluid flow, drastically reducing the efficiency.  In contrast, the flow in the grooved designs look very similar at high and low frequencies.  Note that this may help to reduce the poorly understood flow reversal which can sometimes be observed in plated designs~\cite{urbanski2006,urbanski2007}.

This difference in the sharpness of the performance curves suggests that the
grooved design is more robust and may be less sensitive to geometric or
electrical changes in operational systems.  This is especially important since
the physics behind ACEO is not completely understood, so our theoretical
predictions for optimal geometries and driving frequencies need to be checked
experimentally as in~\cite{urbanski2007}.

\section{Conclusion}
We have predicted that altering the fabrication process for 3D ACEO pumps can
have a major effect on the fluid-conveyor-belt mechanism and the driving slip
velocities.  In particular, designs whose 3D features have non-polarizable
vertical sides should outperform those with polarizable side walls.  We have
explored this hypothesis numerically and have found that the grooved design
could potentially double the pumping velocity of existing plated devices.  The
improved performance comes without having to increase the applied voltages,
while at the same time being more robust to engineering and prediction errors.
Our design principle could therefore have a significant impact on various
lab-on-a-chip applications.

\begin{acknowledgments}
This research was supported by the U.S.\ Army through the Institute for Soldier
Nanotechnologies, under Contract DAAD-19-02-0002 with the U.S.\ Army Research
Office.
\end{acknowledgments}

\bibliography{grooved}

\end{document}